\documentclass[reprint,
groupedaddress,
showpacs,preprintnumbers,
 amsmath,amssymb,
 aps,
prl
]{revtex4-1}
\usepackage{graphicx}% Include figure files
\usepackage{dcolumn}% Align table columns on decimal point
\usepackage{bm}% bold math
\usepackage{mathrsfs}
\begin{document}

%\preprint{}

\title{The microscopic basis for phase-sensitive experiments for determination of the order parameter symmetry in Fe-based superconductors.}
\author{A.\,V.~Burmistrova}
%\thanks{}
\affiliation{Lomonosov Moscow State University Skobeltsyn Institute of Nuclear Physics, 1(2), Leninskie gory, GSP-1, Moscow 119991, Russian Federation}

\author{ I.\,A.~Devyatov}
\email[]{igor-devyatov@yandex.ru}
%\thanks{}
\affiliation{Lomonosov Moscow State University Skobeltsyn Institute of Nuclear Physics, 1(2),  Leninskie gory, GSP-1, Moscow 119991, Russian Federation}

%\author{Alexander A. Golubov}
%\affiliation{Faculty of Science and Technology and MESA+ Institute of Nanotechnology,
%University of Twente, 7500 AE, Enschede, The Netherlands}

%\author{Keiji Yada}
%\affiliation{Department of Applied Physics, Nagoya University, Nagoya 464-8603, Japan}

%\author{Yukio Tanaka}
%\affiliation{Department of Applied Physics, Nagoya University, Nagoya 464-8603, Japan}


\date{\today}

\begin{abstract}
We present a microscopic theory of dc Josephson current, based on the construction of a coherent temperature Green's function in the tight-binding approximation, in junctions with multiband superconductors. This theory is applied to the junctions with multiband Fe-based superconductors (FeBS) described by $s_{\pm}$-wave and $s_{++}$-wave order parameter symmetries, which probably realized in FeBS. We confirm microscopically the previously suggested crucial experiment for determination of the type of the order parameter symmetry in FeBS.
\end{abstract}

\pacs{74.20.Rp,74.70.Xa,74.45.+c,74.50.+r,74.55.+v}
%\keywords{}

\maketitle


%\section{2D model for the contact between s-wave superconductor and superconducting pnictide}{\label{sec3}}

Determination of the symmetry of the order parameter of a new unconventional superconductor is one of the first tasks after their discovery. It is known that the crucial experiments to determine the symmetry of the order parameter in unusual superconductors are the phase-coherent tunneling experiments. %These experiments may be an experiments to study current-voltage characteristics of a junction of a normal metal with a new superconductor or experiment on the Josephson tunneling in superconducting junction  with a new superconductor.

It should be noted that many of the new unconventional superconductors, such as $Sr_2RuO_4$, FeBS, doped superconducting insulators  $Cu_xBi_2Se_3$ are multiorbital metals. Therefore, a quantitative microscopic theory that describes the coherent tunneling in the junctions containing these unconventional superconductors should  take into account both interband  and intervalley scattering at the boundaries. Such a microscopic theory to describe the current of single-particle excitations in junctions  of a normal metal with a multiband superconductor has been proposed only recently \cite{new}.  The aim of this work is the creation of the consistent microscopic theory of the Josephson tunneling in junctions with multiband superconductors and application of this theory to FeBS with the most popular types of the symmetries of the order parameter.  Previous theories devoted to the study of Josephson tunneling in junctions with multiband superconductors are phenomenological \cite{berg,chen,koshelev2012,lind}. Our theory is applied to  calculation of the Josephson current-phase dependencies in junctions between single-band superconductor and FeBS with $s_{\pm}$-wave and $s_{++}$-wave order parameter symmetries for different directions of current with respect to the crystallographic axes of FeBS. Based on these calculations, we demonstrate the possibility to determine the symmetry of the order parameter in FeBS and for the first time microscopically confirm one of the recently proposed for this purpose experimental scheme.

\begin{figure}[h]
\centerline{\includegraphics[width=8cm]{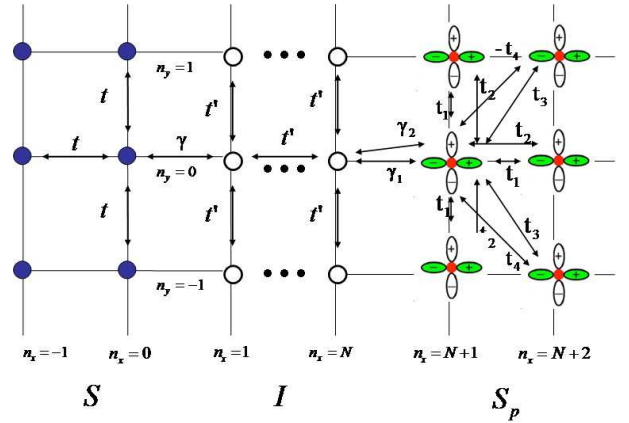}}
\caption{Schematic illustration of $2D$ model of the s-wave superconductor/insulator/FeBS structure. }
\label{SISp}
\end{figure}

We consider a model clean planar superconducting $S/I/S_p$ junction with perfectly flat interfaces in the tight-binding approximation like the one depicted in Fig. \ref{SISp}.  One can see a two-dimensional crystallographic plane of a conventional spin-singlet s-wave superconductor $S$ (blue filled circles on left side of Fig. \ref{SISp}),
N atomic layers of an insulator (circles in the middle of Fig. \ref{SISp})
and multi-orbital superconductor $S_p$ (in the right part of Fig. \ref{SISp}). In Fig. \ref{SISp} $t$ and $t'$ are the hopping parametrs in S and I, respectively.
We consider the application of our method for the case when a multiorbital superconductor $S_p$ is FeBS.

The minimal model to reproduce  Fermi surfaces in FeBS is a two-band model considering $d_{xz}$ and $d_{yz}$ orbitals in iron \cite{rag}.
There are four hopping parameters $t_1$, $t_2$, $t_3$ and $t_4$ in this model, as shown in Fig. \ref{SISp}. For the pair potential, the intra-orbital $s_{\pm}$
and  $s_{++}$ models  are  considered \cite{maz1}.
%These pair potentials correspond to intra-orbital pairing and do not depend on the type of orbital.
We consider  the case of zero misorientation angle of the crystallographic axes of  FeBs with respect to the interface as shown in Fig. \ref{SISp}.
The  hopping parameter between the sites of a usual superconductor $S$ on the left side and the sites of an insulator $I$ is described by $\gamma$, and the  hopping parameters between the sites of an insulator $I$ and $d_{xz}$ ($d_{yz}$)-orbitals
of FeBS are described by $\gamma_1$ ($\gamma_2$).
For simplicity, we assume that
the periods of the crystal lattices in a normal metal and FeBs are the same and equal to $a=1$.
To calculate  the Josephson current across $S/I/S_p$ junction we should construct coherent Green's function of the whole  system.
The simplest way to do it is to construct the Green's functions of  $S,~I,~S_p$ regions and then match them at the boundaries.
Let us define the temperature matrix Green's function $\mathbf{G}$ in the tight-binding approximation for FeBS in the framework of the two-orbital model in the following  form:
\begin{equation}
\begin{aligned}
&\mathbf{G}_{\{n\},\{j\}}(\tau_1, \tau_2)=\begin{pmatrix} \widehat{G}_{\{n\},\{j\}}(\tau_1, \tau_2) & \widehat{F}_{\{n\},\{j\}}(\tau_1, \tau_2)\\ \widehat{\widetilde{G}}_{\{n\},\{j\}}(\tau_1, \tau_2) & \widehat{\widetilde{F}}_{\{n\},\{j\}}(\tau_1, \tau_2)\end{pmatrix},
\end{aligned}\label{green1}
\end{equation}

\noindent where
$\widehat{G}_{\{n\},\{j\}}(\tau_1, \tau_2),\widehat{F}_{\{n\},\{j\}}(\tau_1, \tau_2), \widehat{\widetilde{G}}_{\{n\},\{j\}}(\tau_1, \tau_2)$, $\widehat{\widetilde{F}}_{\{n\},\{j\}}(\tau_1, \tau_2)$ are $4\times4$ matrices in orbital space, which we describe by upper indexes $(\alpha\beta)$:

\begin{equation}
\left\{
\begin{aligned}
&{G}^{(\alpha\beta)}_{\{n\},\{j\}}(\tau_1, \tau_2) =-\langle T_{\tau}c^{(\alpha)}_{\uparrow}(\{n\},\tau_1){c^{(\beta)+}_{\uparrow}}(\{j\},\tau_2)\rangle,\\
&{F}^{(\alpha\beta)}_{\{n\},\{j\}}(\tau_1, \tau_2) =\langle T_{\tau}{c^{(\alpha)+}_{\downarrow}}(\{n\},\tau_1)c^{(\beta)+}_{\uparrow}(\{j\},\tau_2)\rangle,\\
&{\widetilde{G}}^{(\alpha\beta)}_{\{n\},\{j\}}(\tau_1, \tau_2) =-\langle T_{\tau}c^{(\alpha)+}_{\downarrow}(\{n\},\tau_1)c^{(\beta)}_{\downarrow}(\{j\},\tau_2)\rangle,\\
&{\widetilde{F}}^{(\alpha\beta)}_{\{n\},\{j\}}(\tau_1, \tau_2) =\langle T_{\tau}c^{(\alpha)}_{\uparrow}(\{n\},\tau_1)c^{(\beta)}_{\downarrow}(\{j\},\tau_2)\rangle.
\end{aligned}
\right.\label{green2}
\end{equation}

\noindent In Eqs. (\ref{green1},\ref{green2}) indices $\alpha$ and $\beta$ run through all values $1,2$, where index $1$ corresponds to the $d_{xz}$ orbital and index $2$ corresponds to the $d_{yz}$ orbital; $c^{(1)+}_{\sigma}(\{n\},\tau_i) \left(c^{(2)+}_{\sigma}(\{n\},\tau_i))\right)$ is creation  operator  of an electron belonging to the $d_{xz}$ ($d_{yz}$) orbital with spin $\sigma$ on $\{n\}=(n_x,n_y)$ site, $\tau_i$ is an imaginary "time",
and  $T_{\tau}$ is an imaginary "time" ordering operator.

Green's functions of a conventional superconductor $\mathbf{G^S}$ and an insulator $\mathbf{G^I}$  have the same form as in Eqs. (\ref{green1},\ref{green2}), but
without the upper orbital indices.

Gorkov's equations in the discrete case for arbitrary model of the intraorbital superconducting pairing have the following form:
\begin{widetext}
\begin{equation}
\left\{
\begin{aligned}
&(i\omega_m +\mu)G^{(\alpha\alpha),\omega}_{\{n\},\{j\}}-\sum_{\{l\}}{t^{(\alpha)}_{\{n\},\{l\}}G^{(\alpha\alpha),\omega}_{\{l\},\{j\}}}-\sum_{\{l\}}{t^{(\alpha\beta)}_{\{n\},\{l\}}G^{(\alpha\beta),\omega}_{\{l\},\{j\}}}+\sum_{\{l\}}\Delta_{\{n\},\{l\}}F_{\{l\},\{j\}}^{(\alpha\alpha),\omega}=\delta_{\{n\},\{j\}},\\
&(i\omega_m +\mu)G^{(\alpha\beta),\omega}_{\{n\},\{j\}}-\sum_{\{l\}}{t^{(\beta)}_{\{n\},\{l\}}G^{(\alpha\beta),\omega}_{\{l\},\{j\}}}-\sum_{\{l\}}{t^{(\beta\alpha)}_{\{n\},\{l\}}G^{(\alpha\alpha),\omega}_{\{l\},\{j\}}}+\sum_{\{l\}}\Delta_{\{n\},\{l\}}F_{\{l\},\{j\}}^{(\alpha\beta),\omega}=0,\\
&(i\omega_m -\mu)F^{(\alpha\alpha),\omega}_{\{n\},\{j\}}+\sum_{\{l\}}{t^{(\alpha)}_{\{n\},\{l\}}F^{(\alpha\alpha),\omega}_{\{l\},\{j\}}}+\sum_{\{l\}}{t^{(\alpha\beta)}_{\{n\},\{l\}}F^{(\alpha\beta),\omega}_{\{l\},\{j\}}}+\sum_{\{l\}}\Delta^*_{\{n\},\{l\}}G_{\{l\},\{j\}}^{(\alpha\alpha),\omega}=0,\\
&(i\omega_m -\mu)F^{(\alpha\beta),\omega}_{\{n\},\{j\}}+\sum_{\{l\}}{t^{(\beta)}_{\{n\},\{l\}}F^{(\alpha\beta),\omega}_{\{l\},\{j\}}}+\sum_{\{l\}}{t^{(\beta\alpha)}_{\{n\},\{l\}}F^{(\alpha\alpha),\omega}_{\{l\},\{j\}}}+\sum_{\{l\}}\Delta^*_{\{n\},\{l\}}G_{\{l\},\{j\}}^{(\alpha\beta),\omega}=0.
\end{aligned}
\right.\label{G_eq_p}
\end{equation}
\end{widetext}

\noindent In Eq. (\ref{G_eq_p}) $\alpha \neq \beta$,  $t^{(1)}_{\{n\},\{l\}}(t^{(2)}_{\{n\},\{l\}})$ are the hopping parameters between the same $d_{xz}(d_{yz})$ orbitals,  and $t^{(12)}_{\{n\},\{l\}}(t^{(21)}_{\{n\},\{l\}})$ are the hopping parameters between the different orbitals, $\omega_m=\pi T (2m+1)$, $m$ is integer value, $T$ is the temperature.

Discrete Gorkov's equations for the Green's function of a conventional superconductor $\mathbf{G^S}$ and an insulator $\mathbf{G^I}$  have the same form as in Eq. (\ref{G_eq_p}), but
without the  orbital indices, third and fourth term in the right side of Eq. (\ref{G_eq_p}) for $\mathbf{G^I}$ and without third  term in the right side of Eq. (\ref{G_eq_p}) for  $\mathbf{G^S}$. It can be shown that in order to calculate the  Josephson current in the structure under consideration it is enough to solve only Eq. (\ref{G_eq_p}) with $\alpha=1$ or $\alpha=2$, because the remaining system of the equations gives the same results.

To construct the coherent Green's function of the whole $S/I/S_p$ junction one should match Green's functions of $S$, $I$ and $S_p$ regions at the boundaries.
The boundary conditions for matching of wave functions in multiorbital metals  were proposed  in \cite{new}. For temperature Green's functions these boundary conditions  for the case $\alpha=1,~\beta=2$ have the form:

\begin{equation}
\left\{
\begin{aligned}
&tG^{S}_{1,j}=\gamma G^{I}_{1,j},\\
&tF^{S_L}_{1,j}=\gamma F^{I}_{1,j},\\
&\gamma G^{S}_{0,j}=t'G^{I}_{0,j},\\
&\gamma F^{S}_{0,j}=t'F^{I}_{0,j},
\end{aligned}
\right.\label{bc1}
\end{equation}

\begin{equation}
\left\{
\begin{aligned}
&t_1G^{(\alpha\alpha)}_{N,j}+ 2t_3\cos{k_y}G^{(\alpha\alpha)}_{N,j} + 2it_4\sin{k_y}G^{(\alpha\beta)}_{N,j}=\gamma_1 G^{I}_{N,j},\\
&t_1F^{(\alpha\alpha)}_{N,j}+ 2t_3\cos{k_y}F^{(\alpha\alpha)}_{N,j} + 2it_4\sin{k_y}F^{(\alpha\beta)}_{N,j}=\gamma_1 F^{I}_{N,j},\\
&t_2G^{(\alpha\beta)}_{N,j}+ 2t_3\cos{k_y}G^{(\alpha\beta)}_{N,j} + 2it_4\sin{k_y}G^{(\alpha\alpha)}_{N,j}=\gamma_2 G^{I}_{N,j},\\
&t_2F^{(\alpha\beta)}_{N,j}+ 2t_3\cos{k_y}F^{(\alpha\beta)}_{N,j} + 2it_4\sin{k_y}F^{(\alpha\alpha)}_{N,j}=\gamma_2 F^{I}_{N,j},\\
&\gamma_1 G^{(\alpha\alpha)}_{N+1,j}+\gamma_2 G^{(\alpha\beta)}_{N+1,j}=tG^{I}_{N+1,j},\\
&\gamma_1 F^{(\alpha\alpha)}_{N+1,j}+\gamma_2 F^{(\alpha\beta)}_{N+1,j}=tF^{I}_{N+1,j}.
\end{aligned}
\right.\label{bc2}
\end{equation}

\noindent Due to the translational invariance of the structure in the direction parallel to the interface $k_y$ component of the quasimomentum is conserved and the subscripts corresponding to the coordinate of a site in this direction is omitted. We neglect the self-consistency of the pair potential at the outlined above procedure of construction of the coherent Green's function of $S/I/S_p$ junction since, as it was shown in \cite{tanaka97}, it is allowed  in theoretical investigation of the Josephson current in junctions with unconventional superconductors .

The  Josephson current is given by
%as a function of phase difference $\varphi=\varphi_R-\varphi_L$, where $\varphi_R(\varphi_L)$- the phase of the order parameter of the right (left) superconductor, using the following equation:

\begin{equation}
\begin{aligned}
I = \frac{eTt}{i\hbar}&\int\sum_{\omega_m}(G^I_{j,j+1}-G^I_{j+1,j}+ \widetilde{G}^I_{j,j+1}-\widetilde{G}^I_{j+1,j})dk_y.
\end{aligned}
\label{current}
\end{equation}
It can be shown that previous relations for the Josephson current in junctions with both conventional and unconventional superconductors \cite{tanaka97} follow from Eqs. (\ref{G_eq_p})-(\ref{current}).
%After averaging over all possible values of $k_y$ phase dependencies of  Josephson current depend on the following factors:
%1. the values of the hopping parameters at the interface $\gamma_1$ and $\gamma_2$;
%2. the size of the Fermi surface in the s-wave superconductor, because for different sizes of the Fermi surface of S we capture regions with different values of $k_y$ in pnictide and, consequently, different phase dependencies of the Josephson current;
%3. the length of the insulating layer, because increasing the length of this layer leads to the suppresion of the contributions to the average current from regions with large $k_y$.
%4. the ratio of the values of order parameters in $S$ and $S_p$, because for different values of this ratio the ratio between the current contributions to the total current from regions of Fermi surfaces in pnictide with large and small values of $k_y$ will be different and, consiquently, the phase dependence of the total average current can be different.
Eqs. (\ref{G_eq_p})-(\ref{current}) provide the possibility to calculate microscopically the
Josephson current in the $S/I/S_p$ junction for different directions of current relative to the crystallographic axes of FeBS and different symmetries of the order parameter in it.

\begin{figure}
\begin{center}
    \begin{tabular}{cc}
      \resizebox{43mm}{!}{\includegraphics{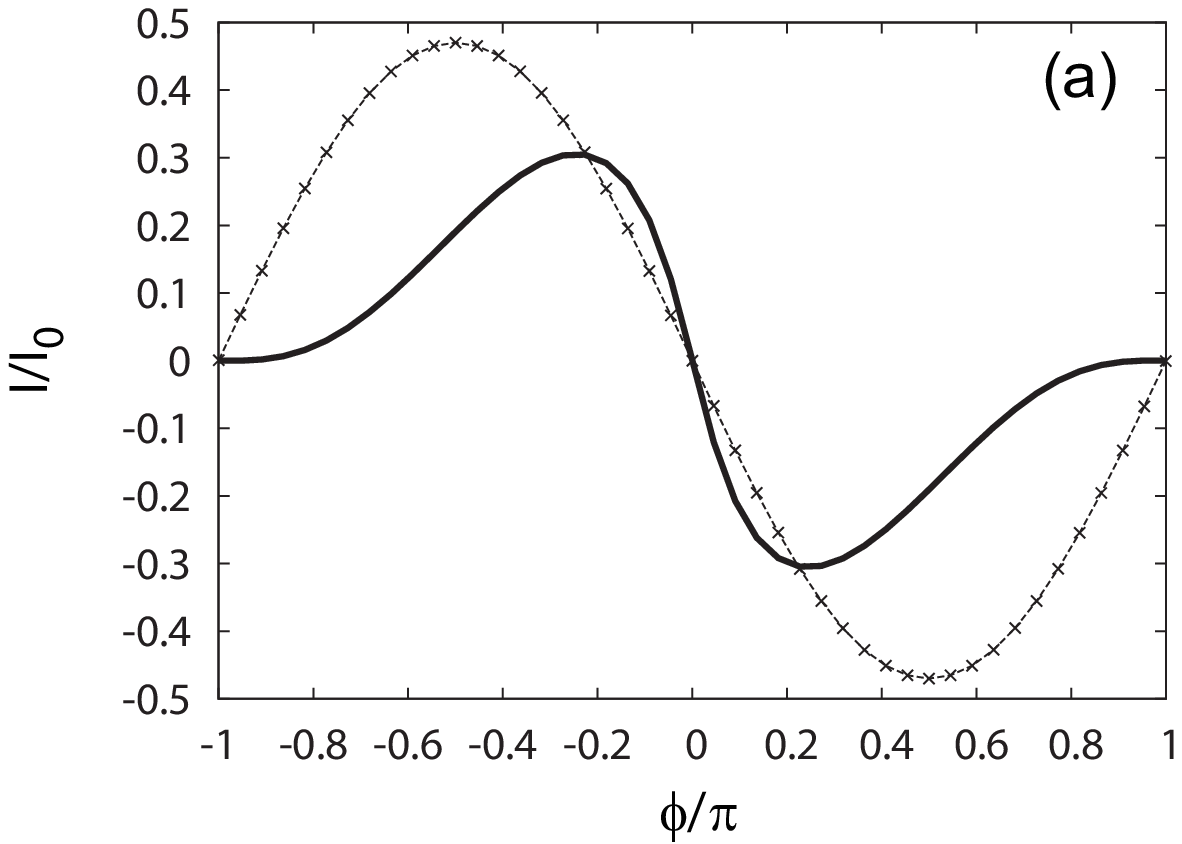}} &
      \resizebox{43mm}{!}{\includegraphics{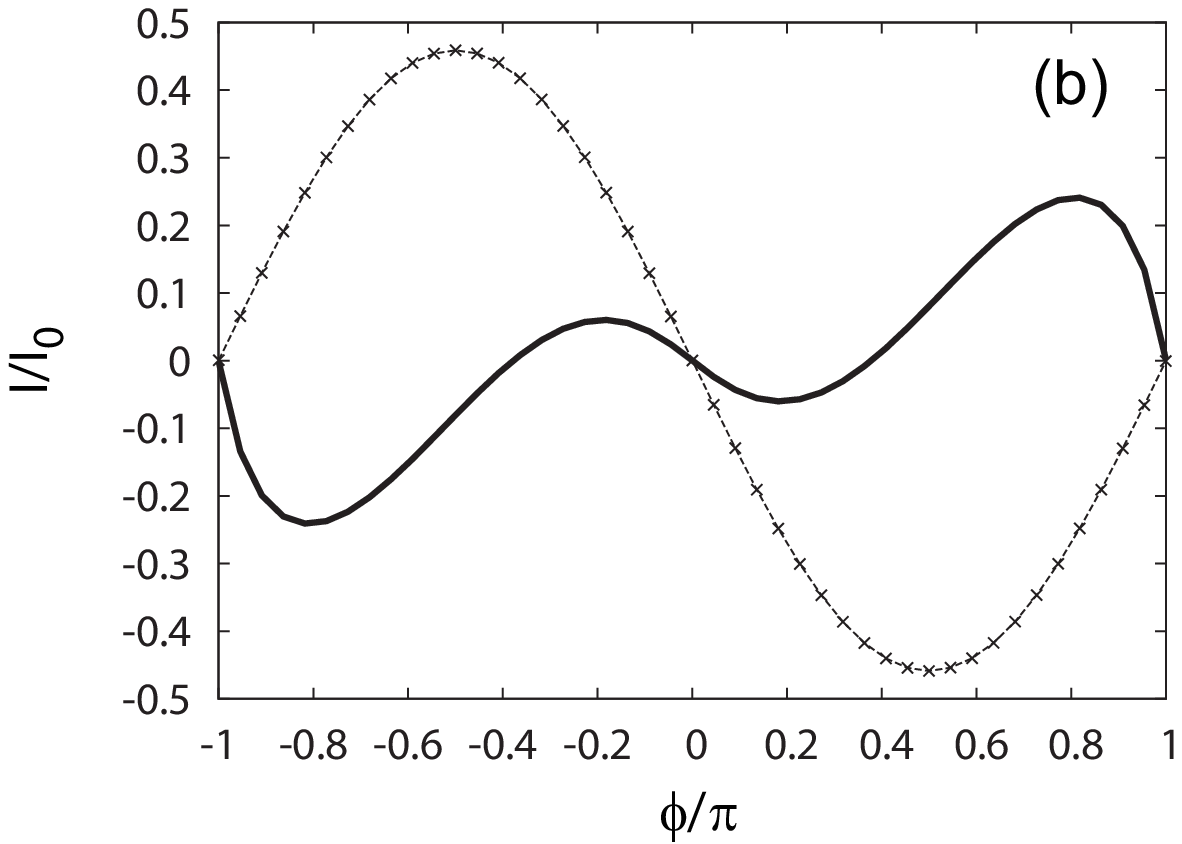}} \\
      \resizebox{43mm}{!}{\includegraphics{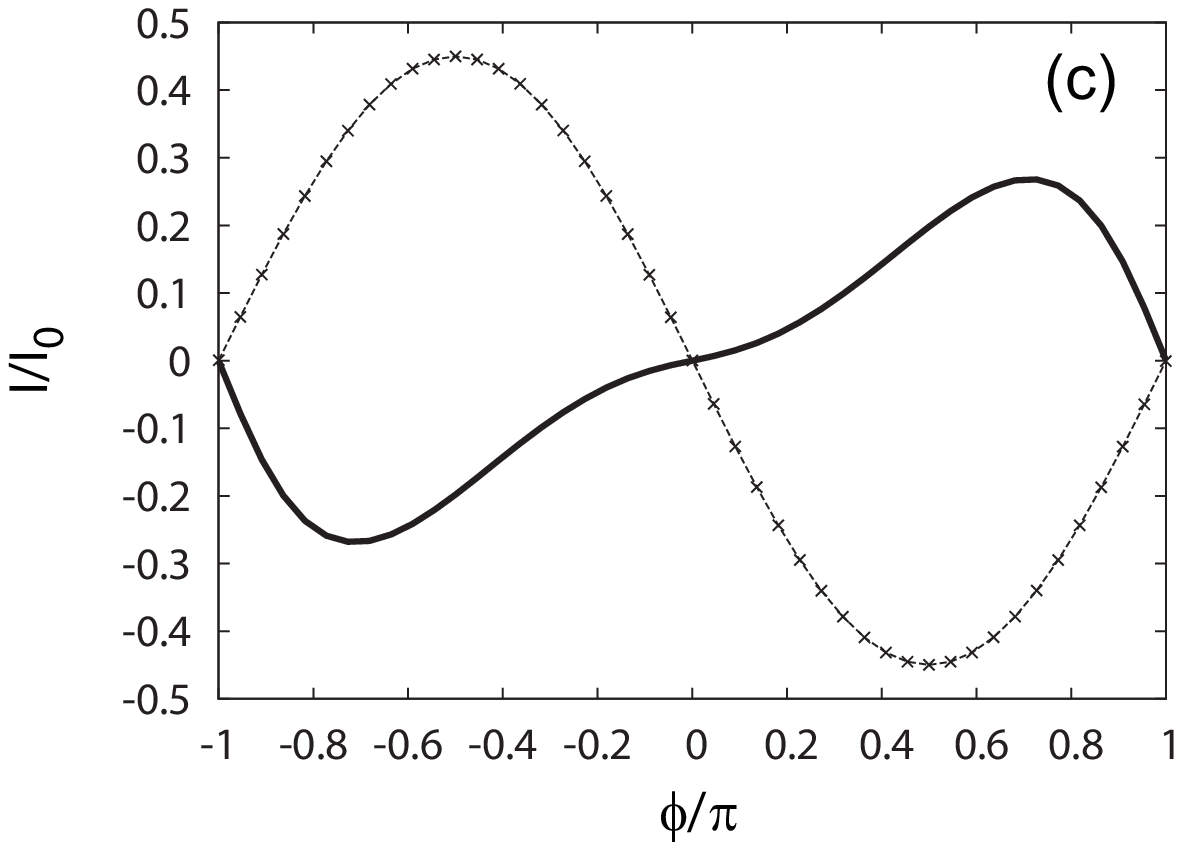}} &
      \resizebox{43mm}{!}{\includegraphics{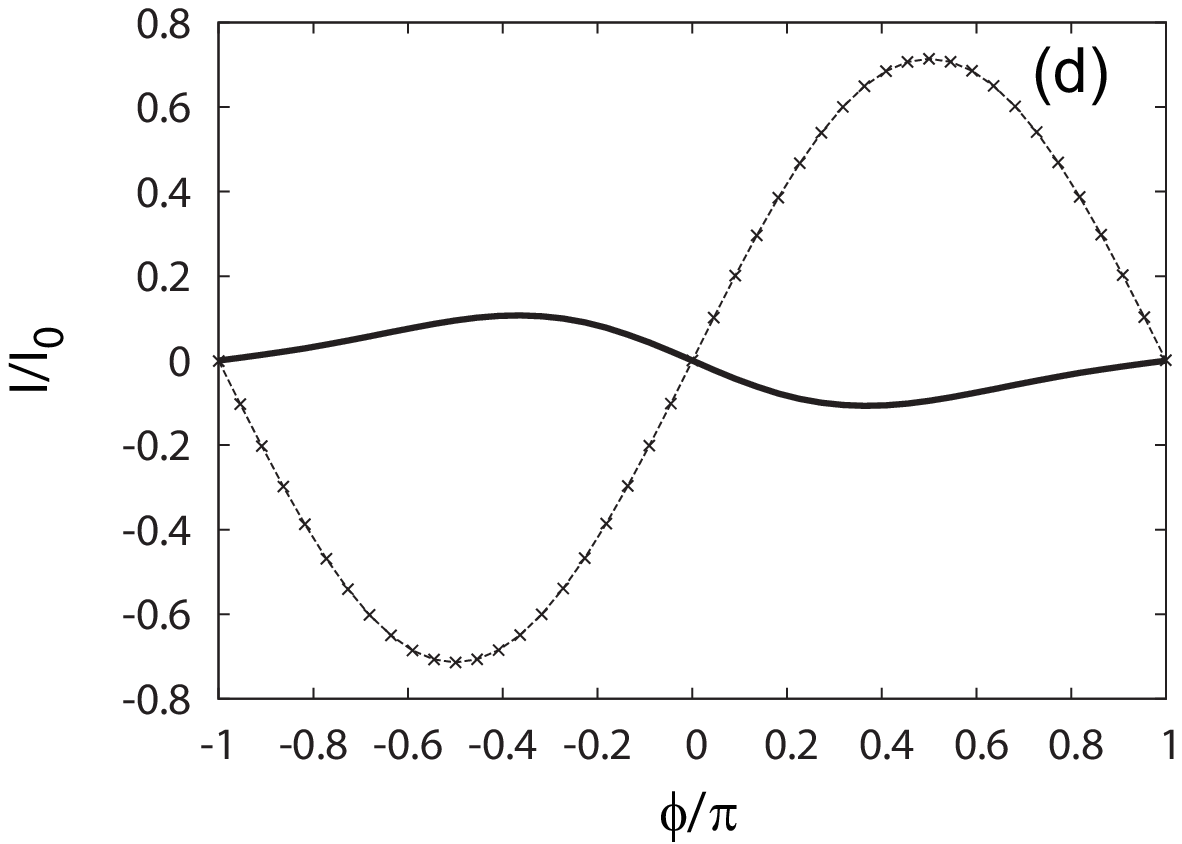}} \\
    \end{tabular}
    \caption{The phase dependencies of the Josephson current in the $S/I/S_p$ junction  for the (100) oriented $S/I/S_p$ junction (Fig. \ref{SISp}), the solid lines correspond to the atomically sharp boundary,  lines with crosses correspond to the $N=3$ layers of an insulator, $I_0=e\Delta_0\sigma_N/\hbar$, where $\sigma_N$-normal conductivity of the structure under consideration; (a) $\gamma_1=0.02, \gamma_2=0.2$; (b) $\gamma_1=0.02, \gamma_2=0.3$; (c) $\gamma_1=0.02, \gamma_2=0.4$; (d) $\gamma_1=0.2, \gamma_2=0.02$.}
    \label{test4}
  \end{center}
  \end{figure}

The phase dependencies of the averaged over  $k_y$ Josephson current in the (100) oriented $S/I/S_p$ junction (Fig. \ref{SISp}) are depicted in Fig.\ref{test4} for the case of the $s_{\pm}$ symmetry of the order parameter in FeBS.
In our calculations we use the following values of  hopping parameters and chemical potential in FeBS:
$t_1 = -0.1051$, $t_2 = 0.1472$, $t_3 = -0.1909$, $t_4 = -0.0874$ and $\mu_{p}=-0.081$ (eV), according to Ref. \cite{mor}, and  suppose that the $S/I$ interface is transparent: $\gamma=t$. We consider
the $s_{\pm}$ model of FeBS with momentum dependent order parameter $\Delta=4\Delta_p\cos k_x\cos k_y$ with $\Delta_p=0.008$ (eV), in a superconductor $S$ we choose the magnitude of the isotropic order parameter $\Delta_0=0.002$ (eV), and suppose relatively low temperature $T/T^s_c \approx 0.02$.
We choose the normal excitation spectrum  in $S$ in the form of $\varepsilon_N=2t(\cos k_x + \cos k_y) + \mu_N$ with hopping parameter $t=-0.3$ (eV) and chemical potential $\mu_N=0.05$ (eV) in order to provide large size of the Fermi surface in $S$. Consequently, areas with large $k_y$ in FeBS contribute to the Josephson current. In the insulating region we choose the normal excitation spectrum in the form of $\varepsilon_I=2t'(\cos k_x + \cos k_y) + \mu_I$ with hopping parameter $t'=-0.3$ (eV) and chemical potential $\mu_I=1.2$ (eV).
In all four figures (a)-(d) in Fig.\ref{test4} the solid lines correspond to the atomically sharp boundary without layers of an insulator,  lines with crosses correspond to $N=3$   layers of an insulator. Figures  (a)-(d) in Fig.\ref{test4} differ from each other by the choice of set of $I/S_p$ interface hopping parameters, which determines the transparency of the $I/S_p$ interface \cite{new}.

One can see from Fig.\ref{test4} that for different sets of $I/S_p$ interface hopping parameters and atomically sharp $I/S_p$ boundary $S/I/S_p$ Josephson  junction can achieve ground state at the phase difference $\varphi=\pi$ (Fig.\ref{test4},a,d), $\varphi=0$ (Fig.\ref{test4},c) and $\varphi=\phi_0$, where $0<\phi_0<\pi$ (Fig.\ref{test4},b).
Such a variety of the current-phase dependencies is explained by  the sign-changing in different bands of the $s_{\pm}$ order parameter in FeBS and the contribution from all values of $k_y$ to the total Josephson current in this case.
Taking into account  of an insulating layer in $S/I/S_p$ Josephson  junction leads to the suppression of the contributions to the average current from regions with large $k_y$, therefore the regions with small $k_y$ dominate \cite{aplmaz}. In this case (lines with crosses in Fig.\ref{test4})
the current-phase dependence becomes very close to the sinusoidal with ground state at $\varphi=\pi$ (Fig.\ref{test4},a,b,c) and $\varphi=0$ (Fig.\ref{test4},d). This situation differs  from the case of $S/I/S_d$ Josephson junctions  with $d$-wave superconductor with nonzero misorientation angle, when $\pi$ - contact survive with increasing the length of an insulator layer \cite{tanaka97}.

Our calculations of the phase dependence of the Josephson current in $S/I/S_p$   junction with the $s_{++}$ symmetry of the order parameter in FeBS demonstrate, that in all cases this junction has ground state at $\varphi=0$.

One can see from the above results that it is difficult to distinguish between the symmetry of the order parameter in FeBS by examining the (100)-oriented $S/I/S_p$    Josephson junction because for different values of the hopping across the $I/S_p$ boundary  $0$, $\pi$ and $\phi$ contacts  can exist, and adding layers of an insulator can lead to the appearance both $0$ and $\pi$ contacts. Also the values of the hopping  parameters across the boundary can be estimated only approximately.

  \begin{figure}
\centerline{\includegraphics[width=8cm]{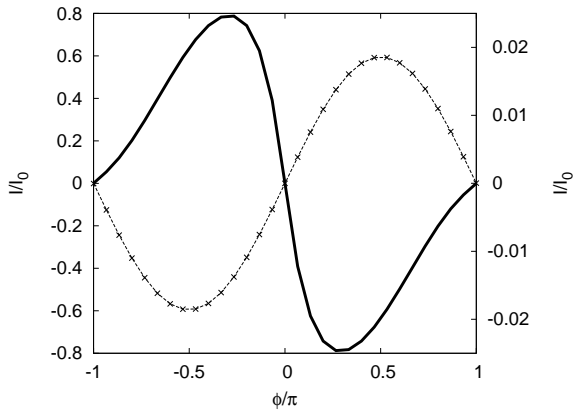}}
\caption{The phase dependencies of the Josephson current in the $S/I/S_p$ junction along $z$-axis.
Solid line and left axis correspond to atomically sharp boundary, line with crosses and right axis correspond to the case of insulating layer containing $N=3$ atoms, $I_0=e\Delta_0\sigma_N/\hbar$. }
\label{R5}
\end{figure}

But the situation changes in the case of investigation of the Josephson current in $S/I/S_p$ junctions along $z$-axis. In this direction at each fixed $k_{||}=(k_x,k_y)$ the contribution to the Josephson current is effectively  just from one of FeBS band, because another band is significantly far from the Fermi level.
The phase dependencies of the averaged over  $k_{||}$ Josephson current along $z$-axis in the $S/I/S_p$ structure are depicted in Fig.\ref{R5}. Solid line and left axis correspond to atomically sharp boundary, line with crosses and right axis correspond to the case  for an insulating layer containing $N=3$ atoms.
In this calculations we choose the normal excitation spectrum in $S$ in the form of $\varepsilon_N=2t(\cos k_x + \cos k_y + \cos k_z) + \mu_N$ with hopping parameter $t=-0.3$ (eV) and chemical potential $\mu_N=0.6$ (eV). Such values of the hopping parameter and chemical potential  provide sufficiently large size of the Fermi surface in $S$, so both electronic and hole packets contribute to the Josephson current. For FeBS along $z$-axis we take into account only hopping  $t_z=-0.1$ (eV) between the same orbitals on the nearest neighbor sites.
We considered transparent $S/I$ interface and the following values for hopping parameters across $I/S_p$ interface: $\gamma_1= \gamma_2=0.17$.
Our calculations demonstrate that in the case of atomically sharp boundaries
the contribution to the total Josephson current from electron pockets dominates and the resulting $S/I/S_p$ junction is $\pi$-junction (solid line in  Fig.\ref{R5}).
The presence of an insulating layer leads to the suppression of the contributions to the averaged current from regions with large $k_{||}$, that is from the electron pockets, so the considered structure with nonzero insulating layer  has ground state at $0$ phase difference (line with crosses in  Fig.\ref{R5}).
It should be noted, that the similar results for the Josephson tunneling in c-direction have been obtained recently using different technique \cite{yada}.
%Modern technology permit to sputter few layers of usual superconductor (for example, Al) and it's further oxidation.
Modern technology permit to create the loop of the normal superconductor, one end of which is oxidized and other is not, connect it with c-oriented FeBS and create dc SQUID.
If one observes in this experiment $\pi$ phase shift, it will be
the crucial evidence in favor of the presence of the $s_{\pm}$ symmetry in FeBS. The same experiment was suggested recently in \cite{aplmaz}.
It is also necessary to note the significant suppression of the magnitude of the Josephson current in the case with long insulator layer compare to atomically sharp boundaries (right and left axis in Fig.\ref{R5}). This results can be one of the explanation of the Josephson critical current suppression in recent Josephson tunneling experiment   in FeBS \cite{siedel}.

In conclusion, we have proposed a microscopic theory describing Josephson tunneling in junctions with multiband superconductors. Our theory takes into account the complex excitation spectrum of these superconductors, their multiband Fermi surface, as well as interband scattering at the boundaries. This theory has been applied to the calculation of the Josephson current-phase relations in junctions of  FeBS described by $s_{\pm}$-wave and $s_{++}$-wave order parameter symmetries with a conventional superconductor for different directions of current relative to the crystallographic axes of FeBS  and different length of an insulator layer.  We have demonstrated the possibility of the ultimate determination of the symmetry of the order parameter in FeBS and for the first time we have confirmed microscopically the recently proposed for this purpose experimental scheme \cite{aplmaz}.

\begin{acknowledgments}
We thank Y. Tanaka, K. Yada, A.A. Golubov  for useful discussions
and acknowledge financial  support from the Russian Foundation for Basic
Research, project N 13-02-01085.
\end{acknowledgments}

%\begin{widetext}

%\end{widetext}

% tables should appear as floats within the text
% The ruledtabular enviroment adds doubled rules to table and sets a
% reasonable default table settings.
% Use the table* environment to get a full-width table in two-column
% Add \usepackage{longtable} and the longtable (or longtable*}
% environment for nicely formatted long tables. Or use the the [H]
% placement option to break a long table (with less control than
% \begin{table}%[H] add [H] placement to break table across pages
% \begin{ruledtabular}
% Lines of table here ending with \\
% \end{tabular}
% \end{ruledtabular}
% \end{table}

% Surround table environment with turnpage environment for landscape
% table
% \begin{table}
% \begin{ruledtabular}
% \end{tabular}
% \end{ruledtabular}
% \end{table}

%\appendix
%\section{}



\bibliography{biblio}

\end{document}